\documentclass[]{pasj01}
\usepackage[T1]{fontenc}
\usepackage{lscape}

\begin{document} 
\Received{}%{yyyy/mm/dd}
\Accepted{}%{yyyy/mm/dd}
%\Published{yyyy/mm/dd}
\title{Detection of the neutral iron line from the supernova remnant W49B with Suzaku}
%%% begin:list of authors
% Do NOT capitalize all letters in "textsc".

\author{
Nari \textsc{Suzuki}\altaffilmark{1,$\ast$}, 
Shigeo \textsc{Yamauchi}\altaffilmark{1}, 
Kumiko K. \textsc{Nobukawa}\altaffilmark{2}, 
Masayoshi \textsc{Nobukawa}\altaffilmark{3}, 
and Satoru \textsc{Katsuda}\altaffilmark{4}
}
\altaffiltext{1}{Department of Physics, Nara Women's University, Kitauoyanishimachi, Nara 630-8506, JAPAN}
\email{wan{\textunderscore}suzuki@cc.nara-wu.ac.jp}
\altaffiltext{2}{Faculty of Science and Engineering, Kindai University, Higashi-Osaka, Osaka 577-8502, JAPAN}
\altaffiltext{3}{Department of Teacher Training and School Education, Nara University of Education, Takabatake-cho, Nara 630-8528, JAPAN}
\altaffiltext{4}{Graduate School of Science and Engineering, Saitama University, 255 Shimo-Ohkubo, Sakura, Saitama 338-8570, JAPAN}
%%% end:list of authors
%% `\KeyWords{}' always has to be placed before `\maketitle'.
 %Do NOT move this preamble from here!
\maketitle

\begin{abstract}
Recent studies of supernova remnants (SNRs) have revealed that 
some SNRs exhibit a neutral iron line emission at 6.4 keV. 
This line has been proposed to originate from the interaction of high-energy particles formed in the SNR shell with the surrounding cold matter. 
We searched for the neutral iron line emission in the SNR W49B. 
Significant detection of the 6.4 keV line is found in the northwest region, close to the molecular cloud interacting with the SNR shell. 
In addition, an excess emission at 8--9 keV, in which K$\gamma$, K$\delta$, and K$\epsilon$ lines of He-like iron exist, 
is also significantly found in the region where the radio shell is not bright.
We discuss the origin of the 6.4 keV line and the excess emission at 8--9 keV.

\end{abstract}
%\pagewiselinenumbers

\section{Introduction}

Supernovae and supernova remnants (SNRs) are major sites for the production of heavy elements and the acceleration of high-energy particles. 
X-ray observations of SNRs are important for performing plasma diagnostics and for the search for X-rays from high-energy particles. 

Cosmic-rays below the knee energy, composed mainly of protons, are thought to be accelerated in the Galaxy. 
SNRs are the prime candidate for origin of the Galactic cosmic-ray. 
High-energy cosmic-ray protons are observed in the gamma-ray band via the $\pi^{0}$ decay process, but
the low-energy cosmic-ray protons (LECRp) are difficult to see. 
When LECRp accelerated in the shell collide with interstellar gas, 
they ionize neutral iron atoms and have them emit the neutral iron line (Fe\,\emissiontype{I} K$\alpha$) at 6.4 keV \citep{Tatischeff2003}. 
Since the 6.4 keV line, which could be due to this phenomenon, has been detected in several SNRs in recent years (e.g., \cite{Nobukawa2018}), 
the 6.4 keV line is a key to understanding cosmic-ray production. 

Recently, survey observations of the 6.4 keV line have been performed, and 
the neutral iron line has been detected from some SNRs (e.g., \cite{Nobukawa2019}). 
The location of the neutral iron line, which correlates with the surrounding molecular cloud, 
suggests that the line originates from the interaction between LECRs and iron atoms in the cold matter.  

%%%%%%
% Figures 1
%%
\begin{figure*}[t]
  \begin{center}
         \includegraphics[width=16cm]{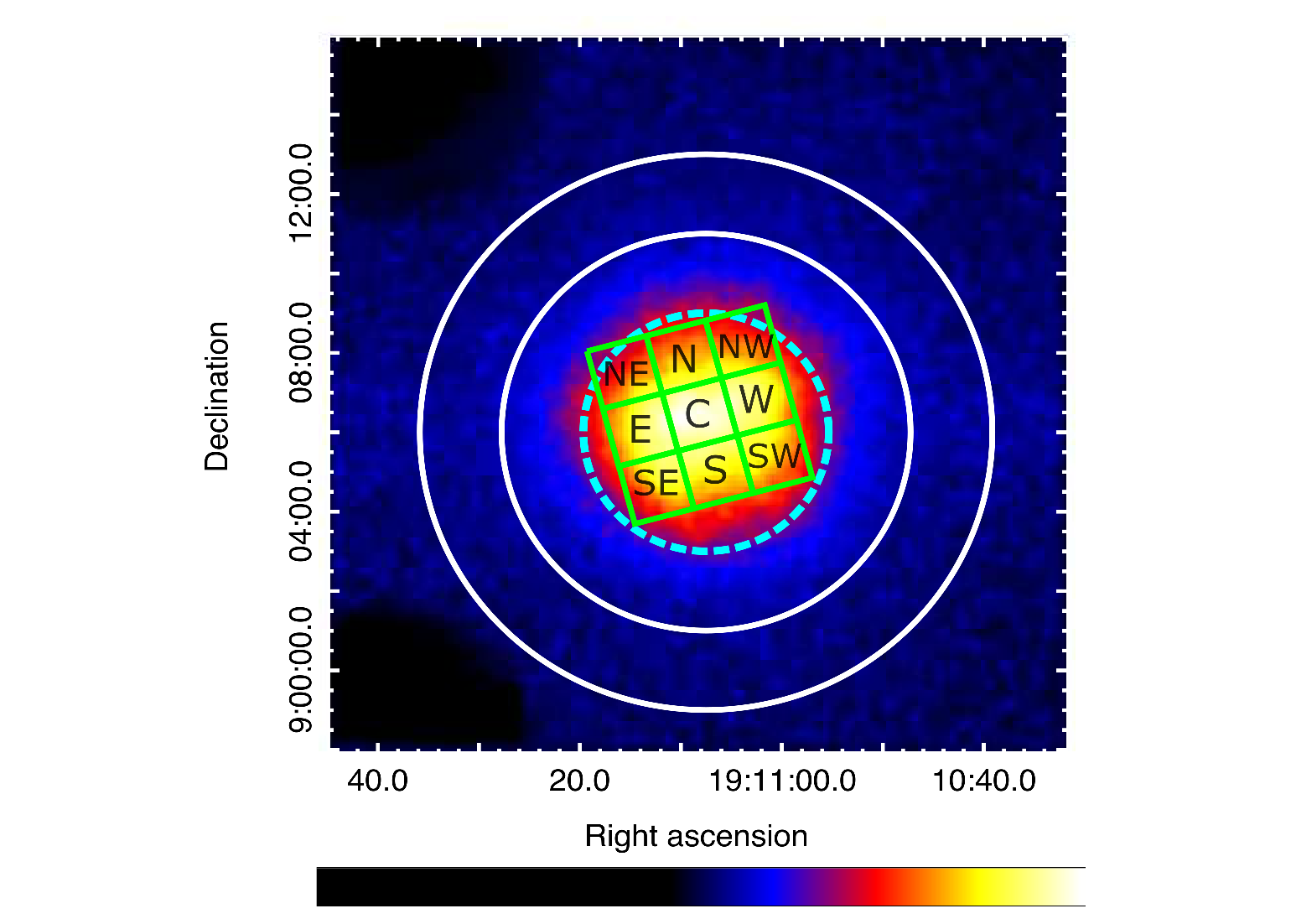}
           \end{center}
  \caption{
XIS image of W49B in the 1--10 keV energy band obtained in 2009.
The coordinates are J2000.0.   
The intensity levels are logarithmically spaced.
The background subtraction and the exposure correction are performed.
The whole region and the 9 regions for the source spectra are shown by a blue dashed line and green solid lines, respectively.
The background region is an annulus shown by white solid lines.
}\label{fig:img}
\end{figure*}
%%%%%%

W49B is an SNR surrounded by molecular clouds. 
Some evidence of interaction between the SNR and the molecular cloud has been reported (e.g., \cite{Kilpatrick2016}; \cite{Sano2021}). 
\citet{Sano2021} reported that the northwestern CO cloud detected by ALMA CO observations interacts with W49B 
because this molecular cloud corresponds well not only to the X-ray structure but also to the radio continuum structure and this molecular cloud has a higher kinetic temperature than the other surrounding clouds.
If LECRp are produced in the shell of W49B, the neutral iron line can be observed from the region where the shell interacts with the molecular cloud. 
In fact, W49B is considered to be an efficient accelerator of cosmic-ray protons (CRp) based on gamma-ray observations (\cite{Abdo2010}; \cite{HESS2018}).
In addition, W49B was found to have an over-ionized plasma (e.g., \cite{Kawasaki2005}; \cite{Ozawa2009}), and there are many highly ionized iron ions.
Therefore, in the iron line band, it is the SNR with the potential for new discoveries not found in standard SNRs.

The instruments onboard Suzaku have better spectral resolution, wider energy band, and lower/more stable intrinsic background than 
previous X-ray satellites \citep{Mitsuda2007},  
and hence they have the best detection efficiency in the iron line band. 
Suzaku observed W49B on several occasions with long exposure times. 
Therefore, we analyzed the Suzaku spectra with high photon statistics and searched for the neutral iron line. 
In this paper, we report on the the 6.4 keV line. 
In addition, an excess emission at 8--9 keV, in which K$\gamma$, K$\delta$, and K$\epsilon$ lines of He-like iron exist, 
is also significantly found in the region where the radio shell is not bright.
The excess component is also discussed.
Throughout this paper, the errors are given at the 90\% confidence level.

\section{Observations and Data Reduction}

%%%%%%%%%%%%%%%%%obs log
\begin{table*}[h]
\begin{center}
  \tbl{Observation logs.}{%
\begin{tabular*}{16cm}{@{\extracolsep{\fill}}llcc@{}}
\hline\noalign{\vskip3pt}
Observation ID&Date start-end&(RA, Dec) [deg]&Exposure [ks]\\ \hline
503084010 &2009-03-29 02:33:12 -- 2009-03-30 11:15:18&(287.7847, 91157)&52.2\\ 
504035010 &2009-03-31 12:43:35 -- 2009-04-02 01:28:20&(287.7847, 91153)&61.4\\
509001010 &2015-04-10 17:55:15 -- 2015-04-13 17:06:24&(287.7875, 9.1067)&113.9\\
509001020&2015-04-13 17:06:25 -- 2015-04-16 16:32:19&(287.7875, 9.1067)&113.2\\
509001030&2015-04-16 16:32:20 -- 2015-04-19 15:53:11&(287.7875, 9.1067)&106.4\\
509001040&2015-04-19 15:53:12 -- 2015-04-21 13:52:15&(287.7875, 9.1067)&67.3\\
\hline\noalign{\vskip3pt}
\end{tabular*}}\label{obs}
\end{center}
\end{table*}
%%%%%%%%%%%%%%%%%%%%%
As shown in table 1,
the Suzaku observations of the SNR W49B were performed 
on 2009 March 29--April 2 (Obs.IDs 503084010 and 504035010) 
and 2015 April 10--21 (Obs.IDs 509001010, 509001020, 509001030, and 509001040) with 
the X-ray Imaging Spectrometer (XIS, \cite{Koyama2007}) placed at the focal planes 
of the thin foil X-ray Telescopes (XRT, \cite{Serlemitsos2007}). 
The XIS was composed of 4 sensors. 
XIS sensor-1 (XIS1) is a back-side illuminated CCD (BI), while
the other three XIS sensors (XIS0, 2, and 3) are a front-side illuminated CCD (FI).
XIS2 turned dysfunctional in 2006 November, 
and a part of the XIS0 was not used due to damage probably caused by the impact of a micrometeorite on 2009 June 23.

The XIS was operated in the normal clocking mode.
The field of view (FOV) of the XIS is \timeform{17.'8}$\times$\timeform{17.'8}.
The XIS employed the spaced-row charge injection (SCI) technique to rejuvenate 
its spectral resolution by filling the charge traps with artificially injected 
electrons through CCD readouts.
Details concerning on the SCI technique are given in \citet{Nakajima2008}
and \citet{Uchiyama2009}.

We retrieved the data processed with the Suzaku pipeline processing version 3 from the DARTS system at ISAS. 
Data reduction and analysis were made using the HEAsoft version 6.30.1 and the calibration database version 2018-10-10.

W49B is a highly absorbed source by the interstellar material.
In this paper, we focus on the iron line band. 
Therefore, the FI data (XIS0 and 3) were used.
The count rates of the XIS0 data obtained in April 2015 were extremely low, only $\sim$60\% of those of the XIS3 data.
This was caused by the telemetry saturation\footnote{$\langle$https://darts.isas.jaxa.jp/astro//suzaku/analysis/doc/suzakumemo/suzakumemo-2015-05.pdf$\rangle$}.
Therefore, the XIS0 data in April 2015 were not used for the present analysis.

\section{Analysis and Results}

Figure \ref{fig:img} shows the X-ray image in the 1--10 keV band. 
We extracted a whole region source spectrum from 
a circular region with a radius of 3$'$, 
while the background spectra were taken from a source free region in the same FOV.
Response files, Redistribution Matrix Files (RMFs)
and Ancillary Response Files (ARFs), were made using
{\tt xisrmfgen} and {\tt xissimarfgen} (\cite{Ishisaki2007}), respectively.

We made background-subtracted spectra according to the following procedure \citep{Hyodo2008}. \\
(1) We constructed the non-X-ray background (NXB) for the source and the background 
spectra from the night earth data using {\tt xisnxbgen} \citep{Tawa2008} and then 
subtracted the NXB from the source and the background spectra. \\
(2) The vignetting effect of the background spectrum was corrected 
by multiplying the effective area ratios between the source and
the background regions for each energy bin.\\
(3) Then, we subtracted the vignetting-corrected background spectra from the source region spectra.

%%%%%%
% Figures 2
%%
\begin{figure}
 \begin{center}
\includegraphics[scale=0.38]{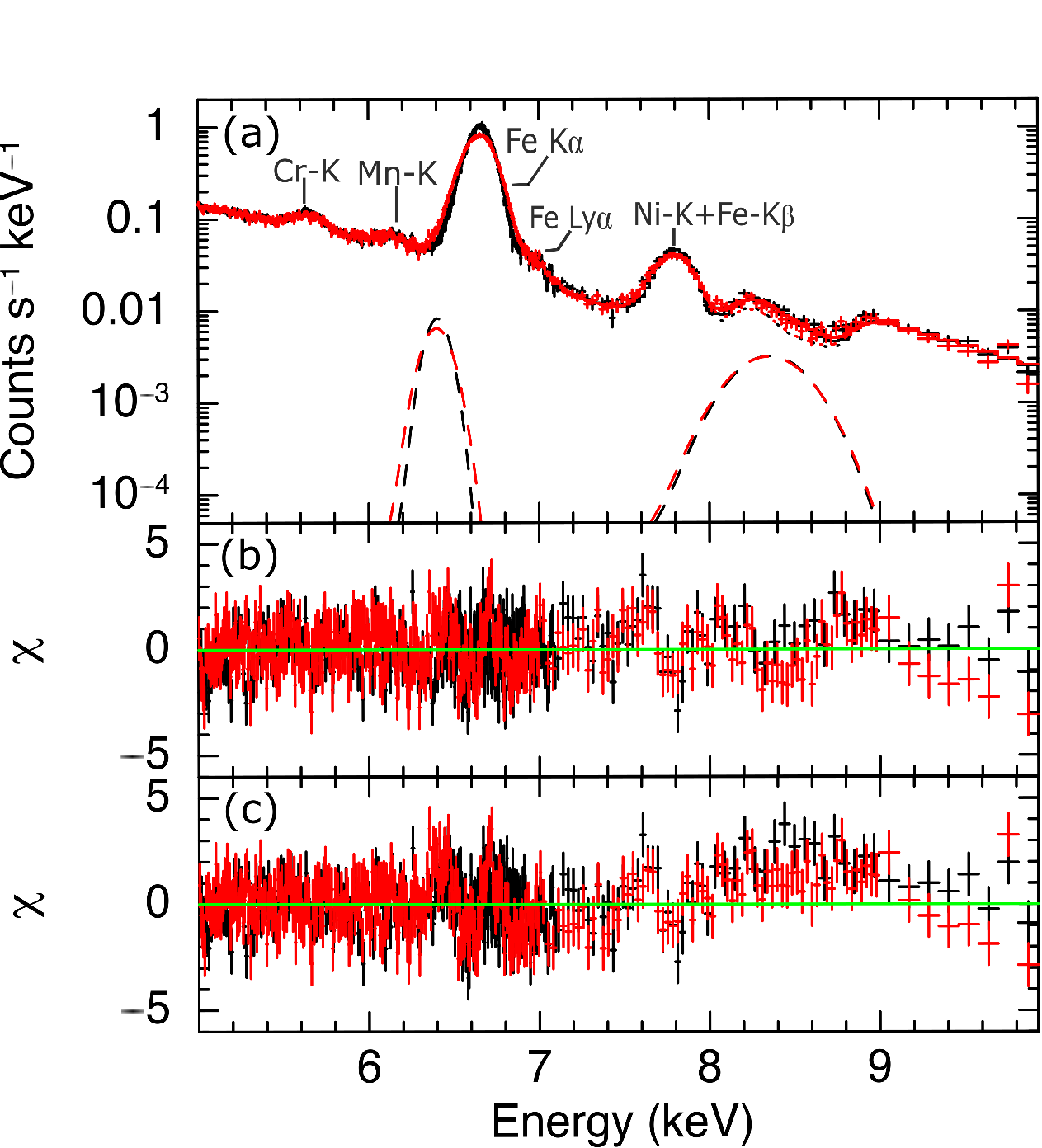}
  \end{center}
 \caption{(a) 
 The whole region spectrum of W49B in the 5--10 keV band.
The crosses show the XIS data. 
The black and red colors show data obtained in 2009 and 2015, respectively.
The histograms show the best-fitting RP$+$2 lines (see table 2).
(b) Residuals from the RP$+$2 lines model. 
(c) The same as (b), but using the best-fitting RP model (see table 2).
}\label{fig:0512spc}
\end{figure}
%%%%%%

%%%%%%
% Table 1
%
\begin{table}[t]
\begin{center}
\caption{The best-fitting parameters for the whole region spectrum.}
\end{center}
\begin{center}
\begin{tabular}{lcc} \hline  
Parameter & \multicolumn{2}{c}{Value} \\
\hline 
Model &    RP   & RP$+$2 lines \\ \hline 
$N_{\rm H}$ ($10^{22}$ cm$^{-2}$)  & 5 (fixed)                            & 5 (fixed)  \\
$kT_{\rm e}$ (keV)                                      & 1.42$\pm{0.01}$    & 1.31$\pm{0.01}$   \\
$kT_{\rm init}$ (keV)                                    &  4.5$\pm{0.1}$        & 5.5 $^{+0.7}_{-0.1}$  \\
$\tau_{\rm RP}^{\ast}$ ($$10$^{11}$ cm$^{-3}$ s) 
                                                                    & 5.0$^{+0.2}_{-0.5}$           & 6.6$^{+1.5}_{-2.3}$   \\
Cr$^{\dag}$                                            & 5.8$\pm{0.3}$              & 6.6 $^{+0.3}_{-0.4}$  \\
Mn$^{\dag}$                                           & 6.5 $^{+0.7}_{-0.8}$     & 8.7  $^{+0.8}_{-0.9}$  \\
Fe$^{\dag}$                                           & 3.2 $\pm$0.1                 & 3.8  $\pm{0.1}$  \\
Ni$^{\dag}$                                             & 8.5 $^{+0.6}_{-0.7}$      & 15.3  $^{+0.8}_{-1.2}$  \\
Normalization$^{\S}$                                 & 0.20 $\pm{0.01}$           & 0.24$\pm{0.01}$      \\ %
$E_{\rm 1}$ (keV)                                       & ---                                 & 6.4 (fixed)  \\
$\sigma_{\rm 1}$ (keV)                                 & ---                                 & 0 (fixed) \\
$I_{\rm 1}$$^{\parallel}$ 
                                                                    & ---                                 & 7.5  $\pm{1.7}$ \\
$E_{\rm 2}$ (keV)                                       &  ---                                 & 8.37$\pm{0.05}$ \\
$\sigma_{\rm 2}$ (keV)                                & ---                                   & 0.20 (fixed) \\
$I_{\rm 2}$$^{\parallel}$ 
                                                                   & ---                                    &  19.6  $\pm{2.3}$  \\ \hline 
$\chi^2$/d.o.f. & 1051.13/645 & 861.04/642  \\
\hline\\
\end{tabular}\label{paraCen}
\end{center}
\vspace{-10pt}
$^{\ast}$ Recombination timescale defined as $\int n_{\rm e} dt$, 
where $n_{\rm e}$ is the electron density (cm$^{-3}$) and $t$ is the time 
(s).\\
$^{\dag}$ Relative to the solar value of Anders and Grevesse (1989).\\
$^{\S}$ Defined as 10$^{-14}$$\times$$\int n_{\rm H} n_{\rm e} dV$ / (4$\pi D^2$),
where $D$ is the distance (cm), $n_{\rm H}$ is the hydrogen density (cm$^{-3}$), 
$n_{\rm e}$ is the electron density (cm$^{-3}$), and $V$ is the volume (cm$^3$). \\
$^{\parallel}$  Units are 10$^{-6}$ photon s$^{-1}$ cm$^{-2}$.\\
\end{table}
%%%%%%

The background-subtracted source spectra in the 5--10 keV band are shown in figure \ref{fig:0512spc}. 
To maximize the photon statistics, the XIS0 and XIS3 spectra obtained in 2009 were merged.
In addition, the XIS3 spectrum obtained in 2015 was fitted simultaneously.
Strong Fe-K (6.68 keV) and Ni-K lines are clearly seen and Cr-K and Mn-K lines are also found. 
In addition, the radiative recombination continuum is clearly found above 9 keV as was reported by \citet{Ozawa2009}.
We fitted the spectra with a recombining plasma (RP) model ({\tt vvrnei} model in XSPEC), modified by a low-energy absorption ({\tt phabs} model in XSPEC).
The free parameters were a current electron temperature ($kT_{\rm e}$), an initial ionization temperature ($kT_{\rm init}$), 
the abundances of Cr, Mn, Fe and Ni, the recombination timescale ($\tau_{\rm RP}$), and the normalization. 
The abundances of the other elements were fixed to be solar values, while the $N_{\rm H}$ value was fixed to $5\times10^{22}$ cm$^{-2}$, 
as was adopted in \citet{Ozawa2009}.
The line and continuum data of the thin thermal plasma were taken from AtomDB 3.0.9. 
The cross section of the photoelectric absorption was taken from \citet{bcmc1992}, while the abundance tables were taken from \citet{Anders1989}.
To adjust the energy scale, we also set the redshift as a free parameter. 
The redshifts were $\sim$10$^{-4}$--10$^{-3}$, which is less than the calibration uncertainty of the energy scale (0.2$\%$ at 6 keV; Suzaku Data Analysis)\footnote{$\langle$http://www.astro.isas.jaxa.jp/suzaku/process/caveats/$\rangle$.}.
This model represented the spectrum with $\chi^2$/d.o.f. $=$ 1051/645 (1.63).
The residuals from the model exhibited a line feature at 6.4 keV. 
We therefore added a gaussian at this energy and obtained an improved fit with $\chi^2$/d.o.f. $=$ 1003/644 (1.56).  
We also fitted the spectrum using only the data from each epoch and evaluated the significance of the 6.4 keV line in each data. 
The decrease of $\chi^2$ in the 2009 and 2015 data were $\Delta \chi ^{2}_{6.4 {\rm keV}}$ ($\Delta$ d.o.f.) = 9.4 (1) and $\Delta \chi ^{2}_{6.4 {\rm keV}}$ ($\Delta$ d.o.f.) = 47.6 (1), respectively. 

%%%8--9 keV
Positive residuals broader than the spectral resolution of the XIS are also found in the 8--9 keV band. 
We fitted it with a broad gaussian. 
The model, with a center energy of 8.37 keV and a line width of 0.2 keV, successfully reproduced the spectrum 
with $\chi^2$/d.o.f. $=$ 861/642 (1.34).
We also fitted the spectrum using only the data from each epoch and evaluated the significance of the broad 8.37 keV line in each data; 
$\Delta \chi ^{2}_{8.37 {\rm keV}}$ ($\Delta$ d.o.f.) = 116.0 (1) for the 2009 data and $\Delta \chi ^{2}_{8.37 {\rm keV}}$ ($\Delta$ d.o.f.) = 81.4 (1) for the 2015 data.
The best-fitting parameters are also listed in table 2, while the best-fitting model is plotted in figure \ref{fig:0512spc}.

In order to examine the spatial variation of the emission line features, 
we extracted source spectra from 9 regions, each \timeform{1.'5}$\times$\timeform{1.'5} square (see figure \ref{fig:img}), 
and fitted the background-subtracted spectra with the same spectral model as that for the whole region spectrum. 
Here, the center energies and the widths of the Gaussians were fixed to the values in table 2. 
The best-fitting parameters are listed in table 3, while the spectra with the best-fitting model are shown in the Appendix.

The electron temperature in the western part is lower than that in the eastern part, 
which is consistent with the NuSTAR results reported by \citet{Yamaguchi2018}. 
On the other hand, although the NuSTAR results showed that the initial ionization temperature is roughly consistent in all the regions \citep{Yamaguchi2018}, it is not the same in our results. 

\begin{figure*}
  \begin{center}
   \FigureFile(8.0cm,8cm){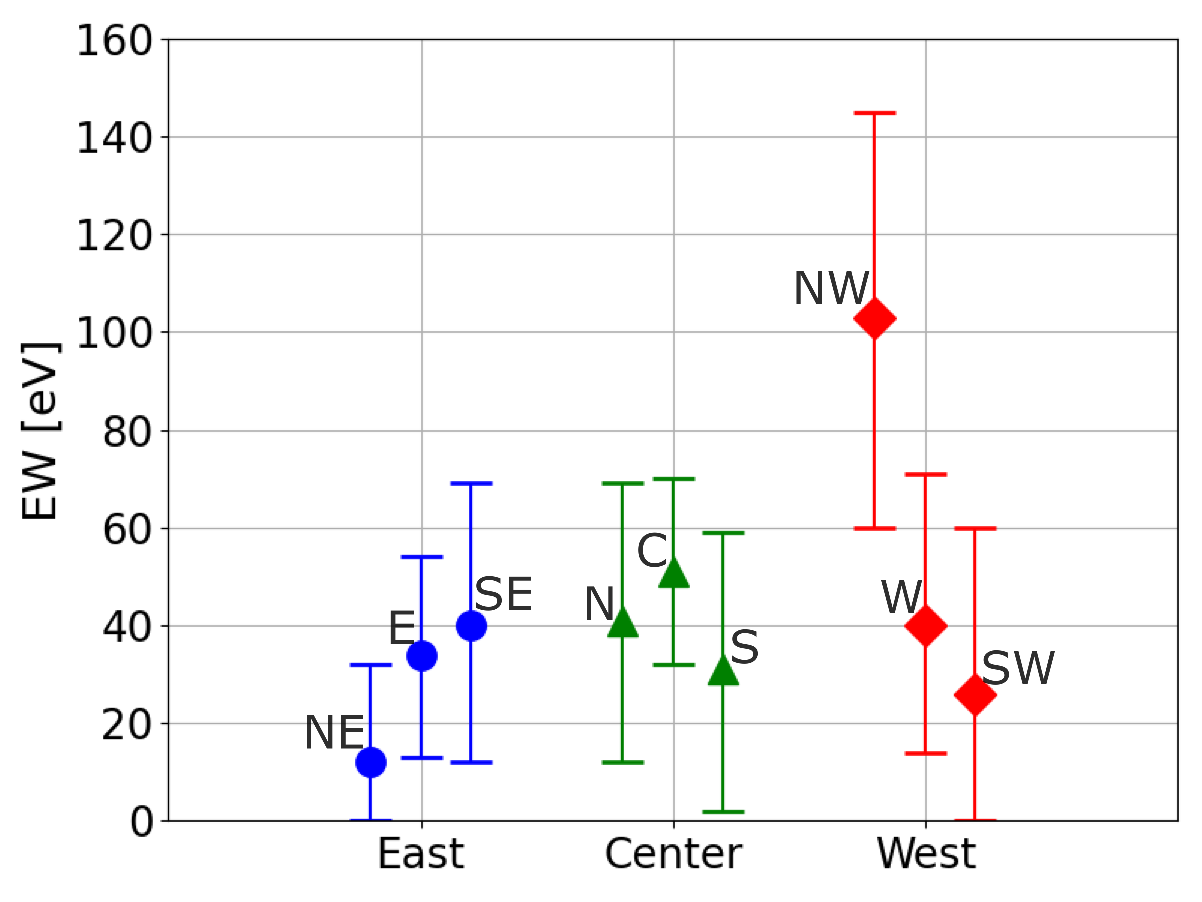}
   \FigureFile(8.0cm,8cm){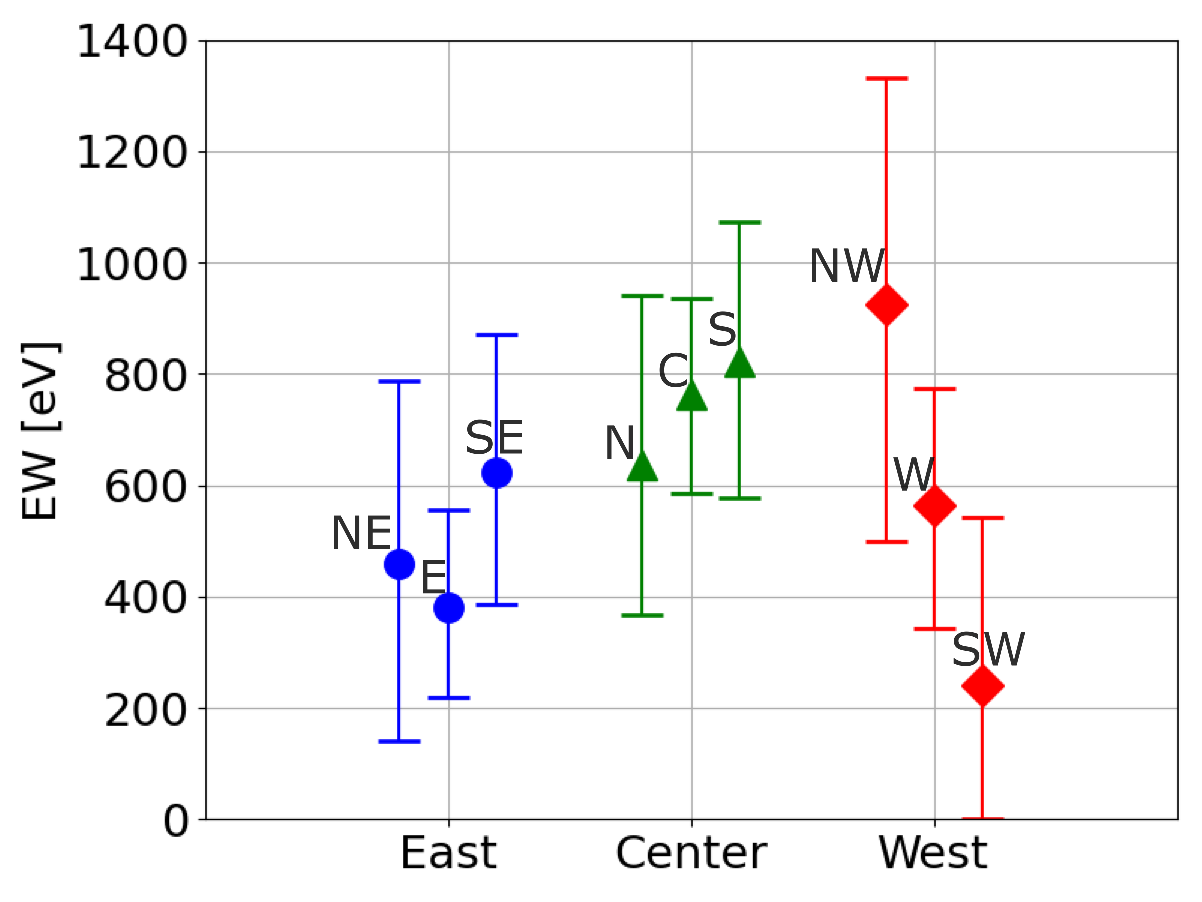}
  \end{center}
  \caption{Left: Equivalent width map of the 6.40 keV line. 
  Right: Same as left, but the broad 8.37 keV line. 
  }\label{fig:EWmap}
\end{figure*}

Figure \ref{fig:EWmap} shows the spatial distribution of the best-fitting equivalent width (EW). 
The EW of the 6.4 keV line is not uniform: the average value for the 8 regions except for the NW region is 34 (9--60) eV, while the value for the NW region is 103 (60--145) eV, indicating a systematic difference from the average.  
On the other hand, that of the excess emission at 8--9 keV is consistent within the 90$\%$ error, but is relatively small in the NE, E, and SW regions.

%%%%%%%%%%%%%%%%%%%%%%%%%%%%%%%%%%%%%%%
%%%%%%
% Table 2
%
\begin{table*}[t]
\caption{The best-fitting parameters for the spatially resolved spectra.}
\begin{center}
\scalebox{0.83}{
\begin{tabular}{lccccccccc} \hline  
Parameter & \multicolumn{9}{c}{Value} \\
Region & NE & E & SE & N & C & S & NW & W & SW\\
\hline 
$N_{\rm H}$$^{\ast, \dag}$  & 5 & 5 & 5 & 5 & 5 & 5 & 5 & 5 & 5 \\
$kT_{\rm e}$ (keV)
  & 1.45$^{+0.08}_{-0.07}$  
  & 1.53$\pm{0.05}$   
  & 1.49$^{+0.09}_{-0.08}$
  & 1.32$^{+0.02}_{-0.06}$
  & 1.32$^{+0.03}_{-0.04}$
  & 1.22$^{+0.06}_{-0.05}$
  & 1.12$^{+0.07}_{-0.08}$
  & 1.13$^{+0.04}_{-0.03}$
  & 1.04$^{+0.07}_{-0.06}$\\
$kT_{\rm init}$ (keV)  
  & 3.6$^{+0.9}_{-0.5}$  
  & $>$ 5.6 
  & $>$ 4.1
  & 7.0$^{+0.3}_{-1.2}$ 
  & 5.5$^{+1.8}_{-1.6}$
  & 5.2$^{+4.3}_{-1.5}$
  & 3.3$^{+1.2}_{-0.6}$
  & 5.5$^{+9.8}_{-0.6}$
  & 2.9$^{+0.3}_{-0.1}$\\
$\tau_{\rm RP}^{\ddag}$ 
  & 4.0$^{+1.9}_{-3.0}$ 
  & 8.5$^{+4.0}_{-2.6}$   
  & 11.1$^{+1.9}_{-6.4}$
  & 7.9$^{+3.2}_{-2.3}$
  & 6.4$^{+1.6}_{-2.5}$
  & 6.5$^{+4.5}_{-2.7}$
  & $<$ 5.3
  & 5.8$^{+2.2}_{-2.1}$
  & $<$ 1.7\\
  %%%%%%%%%%%%%%%Ab
Cr$^{\S}$   
%E 
    & 8.0$^{+1.7}_{-1.8}$ 
    & 7.1$\pm{1.0}$
    &  8.1$^{+1.6}_{-1.5}$
 %C
    &  7.8$\pm{1.3}$
    &  6.3$^{+1.0}_{-0.9}$
    &  5.8$^{+1.0}_{-1.3}$
  %W
    &  4.9$^{+1.7}_{-1.5}$
    &  5.8$^{+0.9}_{-1.0}$
    & 3.4$^{+1.1}_{-0.8}$\\
Mn$^{\S}$   
   & 18.1$^{+4.2}_{-4.3}$
   & 9.0$\pm{2.2}$ 
   &  9.3$^{+3.6}_{-3.3}$
   & 11.0$^{+2.9}_{-3.3}$
   &  9.0$^{+1.9}_{-2.2}$
   & 7.6$^{+2.4}_{-3.0}$
   & 2.9$^{+3.6}_{-2.7}$
   & 6.0$^{+1.8}_{-2.0}$
   & 2.4$^{+0.9}_{-1.6}$\\
Fe$^{\S}$     
%E 
    & 5.0$^{+0.4}_{-0.3}$ 
    & 4.6$\pm{0.2}$
    &  4.0$^{+0.5}_{-0.3}$
 %C
    &  4.3$\pm{0.2}$
    &  4.3$^{+0.2}_{-0.8}$
    &  3.0$\pm{0.5}$
  %W
    &  2.1$^{+0.5}_{-0.7}$
    & 2.2 $\pm{0.3}$
    & 0.8$^{+0.2}_{-0.1}$\\
 Ni$^{\S}$     
%E 
    & 13.0$^{+3.0}_{-3.2}$ 
    & 14.1$^{+3.7}_{-3.1}$
    &  15.6$^{+6.8}_{-4.6}$
 %C
    &  18.4$^{+2.8}_{-3.6}$
    &  18.4$^{+3.6}_{-3.9}$
    &  15.5$^{+6.3}_{-6.7}$
  %W
    &  8.2$^{+7.9}_{-3.3}$
    & 6.5 $^{+2.3}_{-1.6}$
    & 1.0$^{+1.2}_{-0.8}$\\
$E_{\rm 1}$$^{\ast}$ (keV)  & 6.4 & 6.4 & 6.4 & 6.4 & 6.4 & 6.4 & 6.4 & 6.4 & 6.4 \\
$\sigma_{\rm 1}$$^{\ast}$ (keV) & 0 & 0& 0& 0& 0&0 &0 &0 & 0 \\
$I_{\rm 1}$$^{\parallel}$ 
   & 0.05 ($<$ 0.14)
   & 0.63 $^{+0.36}_{-0.39}$ 
   & 0.40 $\pm{0.28}$
   & 0.22 $\pm{0.15}$
   & 1.22 $^{+0.46}_{-0.45}$
   & 0.29 $^{+0.26}_{-0.27}$
   & 0.29 $\pm{0.12}$
   & 0.43 $^{+0.33}_{-0.28}$
   & 0.14 ($<$ 0.32)\\ 
$EW_{\rm 1}$ (eV)
   & 12  ($<$ 32)
   & 34  $^{+20}_{-21}$
   & 40  $^{+29}_{-28}$
   &  41 $^{+28}_{-29}$
   &  51 $\pm{19}$
   &  31 $^{+28}_{-29}$
   &  103 $^{+42}_{-43}$
   &  40 $^{+31}_{-26}$
   & 26 ($<$ 60)\\
%8.37 keV
$E_{\rm 2}$$^{\ast}$( keV) & 8.37 & 8.37 & 8.37 & 8.37 & 8.37 & 8.37 & 8.37 &8.37 & 8.37 \\
$\sigma_{\rm 2}$$^{\ast}$ (keV) & 0.20 & 0.20 & 0.20 & 0.20 & 0.20 & 0.20 & 0.20 & 0.20 & 0.20 \\
$I_{\rm 2}$$^{\parallel}$ 
   & 0.29 $\pm$ 0.20
   & 1.04 $^{+0.48}_{-0.44}$
   & 0.92 $^{+0.35}_{-0.37}$
   & 0.45 $^{+0.22}_{-0.19}$
   & 2.43 $^{+0.55}_{-0.57}$
   & 1.04 $^{+0.32}_{-0.31}$
   & 0.33 $\pm{0.15}$
   & 0.89 $^{+0.34}_{-0.35}$
   & 0.18 ($<$ 0.42)\\
$EW_{\rm 2}$ (eV)
   & 459 $^{+327}_{-318}$
   & 381 $^{+175}_{-162}$
   & 622  $^{+249}_{-236}$
   & 636  $^{+304}_{-269}$
   & 763  $^{+173}_{-178}$
   & 821  $^{+251}_{-245}$
   & 924  $^{+407}_{-426}$
   & 563  $^{+212}_{-219}$
   & 240 ($<$ 542)\\
$\chi^2$/d.o.f. &134.53/126& 274.99/224& 230.89/202 & 236.74/200 & 284.66/222 & 214.17/190 & 164.26/128 & 127.92/128 & 123.33/128 \\
\hline\\
\end{tabular}}\label{para9reg}
\end{center}
\vspace{-10pt}
$^{\ast}$ Fixed value. \\
$^{\dag}$ Units are $10^{22}$ cm$^{-2}$.\\
$^{\ddag}$ Recombination timescale defined as $n_{\rm e} t$ (10$^{11}$ cm$^{-3}$ s), 
where $n_{\rm e}$ is the electron density (cm$^{-3}$) and $t$ is the time 
(s).\\
$^{\S}$ Relative to the solar value of Anders and Grevesse (1989).\\ 
$^{\parallel}$  Units are 10$^{-6}$ photon s$^{-1}$ cm$^{-2}$ arcmin$^{-2}$.\\
\normalsize
\end{table*}

%\end{landscape}
%%%%%%
%%%%%%%%%%%%%%%%%%%%%%%%%%%%%%%%%%%%%%%

%%%%%%
% Figures 3
%%
%%%%%%

\section{Discussion}

%%% 6.4 keV line 
\subsection{6.4 keV line}
Using the Suzaku data in 2009 and the previously unused Suzaku data in 2015 to increase the photon statistics,  
we fitted the X-ray spectrum with a model that includes a 6.4 keV line. 
Since this 6.4 keV line is located at the tail of a strong emission line at 6.7 keV, 
the response indefiniteness is evaluated using a calibration source spectrum, which gives only 0.2$\%$ of the intensity of the line at about 0.3 keV below the K$\alpha$ line.
The 6.4 keV line relative to 6.7 keV in the NW  is 13 times higher than the calculated value, 
while that in the other regions are 2--7 times higher than that.
Thus, this appears to be a source-derived component rather than the indeterminate of the response.

If the 6.4 keV line originates from the SNR plasma, the line intensity should correlate with the continuum intensity, and thus the EW would be constant in all the regions.
The EW of the 6.4 keV line is large in the NW region ($\sim$100 eV), where the dense molecular clouds are located, 
while those in the other regions are $\sim$20--40 eV. 
This suggests that most of the line in the NW region comes from molecular clouds. 

As discussed in \citet{Nobukawa2018}, there is the possibility of ionization of neutral iron by photons from nearby bright sources or cosmic-ray protons/electrons.
Only X-rays above 7.1 keV contribute to the 6.4 keV line production. 
The X-ray luminosity of W49B above 7.1 keV is estimated to be $5.7\times10^{34}$ erg s$^{-1}$, which is only $\sim$4$\%$ of the luminosity required for ionization. 
Therefore, a photoionization scenario with W49B is difficult. 
Since there is no bright light source close enough to explain the intensity of the fluorescence line at 6.4 keV, 
we consider the ionization of neutral iron by LECRp and low-energy cosmic-ray electrons (LECRe).

When LECRp or LECRe collide with interstellar gas, 
they ionize neutral iron atoms and have them emit the neutral iron line at 6.40 keV. 
In SNRs where the 6.4 keV line has been detected, 
the intensities of the 6.4 keV line are $\sim$10$^{-1}$ photons s$^{-1}$ cm$^{-2}$ sr$^{-1}$ (e.g., \cite{Nobukawa2019}).
The intensity obtained in W49B is $I_{\rm{6.4 keV}}$ = 3.4 photons s$^{-1}$ cm$^{-2}$ sr$^{-1}$, which is one order of magnitude larger than the other SNRs.
Since the total energy of the CRp in W49B is larger than that of the other SNRs \citep{Sano2021}, 
the intensity would be larger than the other SNRs because there would be still many LECRp that have been produced and have not escaped or thermalized.

According to \citet{Nobukawa2018}, we roughly calculate 
the total energy of LECRp which are accelerated in the SNR shell and can produce the 6.4 keV line. 
We estimated the energy density of LECRp using 
\begin{equation}
   I_{\rm{6.4 keV}} = \frac{1}{4\pi}N_{\rm H_{\rm MC}} \int \sigma_{\rm{6.4 keV}}\ v\ Z_{\rm Fe}\ A E^{-\gamma}\ dE,
\label{eq1}
\end{equation}
where $\sigma_{\rm{6.4 keV}}$ is the ionization cross section between protons and neutral iron (a peak of $\sim$1.3$\times$10$^{-26}$ cm$^2$ hydrogen-atom$^{-1}$ at 10 MeV, \cite{Tatischeff2012}), 
$v$ is the velocity of the protons, 
$Z_{\rm Fe}$ is the iron abundance (2.1, for the NW region), 
$A$ is the number density at 1 MeV, 
$N_{\rm H_{\rm MC}}$ is the column density of the cold material, 
and $E$ is the energy of the protons. 
Figure 5 of \citet{Tatischeff2012} shows that the cross sections producing 6.4 keV from proton collision has a peak in the MeV band. 
Thus, we calculated the energy density of protons in the MeV band.
Adopting $I_{\rm{6.4 keV}}$ = 3.4 photons s$^{-1}$ cm$^{-2}$ sr$^{-1}$, $\gamma$ = 2, $N_{\rm H_{\rm MC}}$ $\sim$10$^{22}$ cm$^{-2}$, and normalization values for the NW region, 
we obtained the energy density ($ \int AE^{-\gamma+1}dE$) of $\sim$300 eV cm$^{-3}$ in the energy range of 0.1--1000 MeV for protons.
The energy density for protons is much higher than the canonical value of $\sim$ 1 eV cm$^{-3}$ for Galactic cosmic-rays (\cite{Neronov2012}).
There are some observations that the energy density is larger than 1 eV cm$^{-3}$ 
at SNRs from which 6.4 keV lines are detected (e.g. \cite{Nobukawa2018}).
Since the energy density varies with the value of $\gamma$, the present evaluation is not exact, but it is consistent with previous studies (e.g. \cite{Nobukawa2018}).
In the case of electrons, the cross sections producing 6.4 keV from electron collision has a peak in the keV band. 
Adopting the cross sections of $\sim$7.0$\times$10$^{-27}$ cm$^2$ hydrogen-atom$^{-1}$ at 20 keV, 
the same calculation using Eq. (1) yields the energy density of
$\sim$1 eV cm$^{-3}$ in the energy range of 0.1--1000 keV.

The NuSTAR observations (\cite{Tanaka2018}) have detected a power-law-like component consistent with non-thermal bremsstrahlung.  
Therefore, we conducted a fit with the power-law component set at the value established by \citet{Tanaka2018} ($\Gamma$ = 1.4, flux = 3.3$\times$10$^{-13}$ erg cm$^{-2}$ s$^{-1}$ in the 10--20 keV band). 
The flux level of the power-law component 
was more than an order of magnitude smaller than {\tt vvrnei}.  
The EW of the 6.4 keV line was estimated to be larger than 1 keV, on the assumption that this power-law component of $\Gamma$ = 1.4 represents a continuous component.
This value indicates a proton origin (\cite{Dogiel2011}).

%%% 8--9 keV excess
\subsection{Excess emission in the 8--9 keV energy band}

In the single RP model fit, we found systematic residuals in the 8--9 keV band (see  figure \ref{fig:0512spc}). 
Prominent emission lines in the 8--9 keV band are Fe\,\emissiontype{XXV} He$\gamma$ (8.295 keV), Fe\,\emissiontype{XXV} He$\delta$ (8.487 keV), and Fe\,\emissiontype{XXV} He$\epsilon$ (8.588 keV) lines.
The excess in the 8--9 keV band could be the Fe\,\emissiontype{XXV} K complex (n $\geq$ 4 $\rightarrow$ 1).
One possibility is insufficient line intensity due to the recombination process in the current plasma code. 
In this case, the excess emission would correlate with the continuum intensity, and thus the EW would be the same throughout W49B. 
The observed EW of the excess emission is consistent within the 90$\%$ errors but shows variation; 
it is enhanced in the NW region.
There appears to be a direct correlation between EW and the electron temperature: the EW is more pronounced towards the west side, where electron temperature is comparatively lower, than towards the east side, as shown in figure 3.
One may argue that the correlation is due to the lower temperature in the western region, which weakens the strength of the continuum emission. 
However, the data of the three western regions (NW, W, and SW) do not follow the correlation. 
Therefore, it is unlikely that this is the only influence.

Alternatively, the feature is due to the charge exchange (CX). 
In the SNR, CX X-ray emission between highly ionized oxygen and hydrogen atom has been reported in \citet{Katsuda2011} and \citet{Katsuda2012}.
The CX process between highly ionized iron and hydrogen atom in the cold matter would occur as discussed in \citet{Wargelin2005} and \citet{Mullen2016}. 
As shown by the presence of RP, there are many highly ionized iron atoms (Fe\,\emissiontype{XXVI}) in W49B. 
Electrons trapped in higher energy levels of Fe\,\emissiontype{XXVI} transition to the ground state, 
and thus we would observe extra emission lines from Fe\,\emissiontype{XXV} at 8--9 keV (see figure 2 in \cite{Wargelin2005}).  
Here, we discuss the possibility.

We calculate the emissivity of CX in W49B using the following equation, 
\begin{equation}
   P_{\rm{CX}} = \sigma_{\rm{CX}}\ n_{\rm{H}}\ n_{\rm{Fe_{XXVI}}}\ V\ v_{\rm{Fe}},
\label{eq2}
\end{equation}
where $\sigma_{\rm{CX}}$ is the CX cross-section between neutral hydrogen and iron ion ($\sim$10$^{-14}$ cm$^2$, \cite{Mullen2016}), 
$n_{\rm H}$ is the neutral hydrogen density, 
$n_{\rm{Fe_{XXVI}}}$ is the H-like iron ion density, 
V is the CX-emitting volume, 
and $v_{\rm{Fe}}$ is the relative iron and hydrogen ion velocity ($\sim$1000 km s$^{-1}$). 
We assume that $n_{\rm H}$ and $n_{\rm{Fe_{XXVI}}}$ are values comparable to the electron and the iron ion densities calculated from the best-fitting values of the RP component. 
Based on the Suzaku image, we assume that the entire W49B can be represented as a cylinder with a height of approximately 3.0 arcmin and a diameter of 3.9 arcmin, 
which is 8.1 pc and 10.6 pc at the distance of 9.3 kpc (e.g., \cite{Zhu2014}), respectively. 
Then, we calculate the electron and H-like iron ion densities of the RP to be $\sim$4 cm$^{-3}$ and $\sim$10$^{-3}$ cm$^{-3}$, respectively. 
Here, we roughly assume $10\%$ of the ion fraction of Fe\,\emissiontype{XXVI} to the total. 
We utilized the thickness of the CX layer interacting with the molecular cloud of $6.25\times10^{13}$ cm, which is the mean free path of neutral hydrogen moving through the post-shock plasma with a proton density of 16 cm$^{-3}$, 
which is assumed to be four times higher than $n_{\rm H}$ in the RP. 
Here, we took the hydrogen-proton CX cross section of $\sim$10$^{-15}$ cm$^2$ (\cite{Wargelin2008}), which is an order of magnitude smaller than that for the hydrogen-iron ion CX used in Eq.(2). 
The CX-emitting volume facing the west molecular cloud is calculated to be $\sim$10$^{52}$ cm$^3$. 
Based on $P_{\rm{CX}}$ and the fraction of the iron K$\gamma$ emission line emitted in one CX ($\sim$10$\%$, \cite{Mullen2016}), we obtain the CX flux ($P_{\rm{CX}}/4\pi$D$^2$) of $\sim$10$^{-4}$ photons cm$^{-2}$ s$^{-1}$.
This result is well above the observed value of $\sim$10$^{-7}$--10$^{-6}$ photons cm$^{-2}$ s$^{-1}$.
Considering the ionization structure of the post-shock region, the hydrogen density around highly-ionized iron ions would be much lower than the assumed $n_{\rm H}$.
Therefore, the CX flux could be much, say a few orders of magnitude, lower than that estimated above.
Thus, the flux obtained is not inconsistent with the CX scenario. 

The EW of the excess emission is large near the molecular cloud (NW region) and small where it is bright in radio continuum synchrotron observations (NE, E, and SW regions) although the variation is only marginal with the current data. 
This trend is consistent with the results observed in the Cygnus Loop and Puppis A (\cite{Uchida2019}; \cite{Katsuda2012}). 
Thus, in terms of the CX flux and the spatial structure, the extra emission lines from Fe\,\emissiontype{XXV} at 8--9 keV in W49B may be interpreted as the result of the presence of CX emission.
However, in terms of the observed center energy, it is difficult to say that they arise from CX processes. 
The center energy of the 8--9 keV bump is measured to be about 8.37 keV, 
indicating that the principal quantum number ($n$) is 4--5. 
This outcome is incongruous with the experimental and theoretical data (\cite{Wargelin2005}; \cite{Mullen2016}), 
in which $n$ and the line center energy are expected to be 8--9 and $\sim$8.7 keV, respectively, for a reasonable condition of W49B. 

An accurate model of a recombination plasma has not yet been fully established. 
The Suzaku CCD resolution indicates that a broad line at 8--9 keV is complex, thereby restricting any further discussion of it.
It will become clearer what kind of line is originally appearing at 8--9 keV by observations with better resolution, such as XRISM.

\section{Conclusion}
The Suzaku spectrum of W49B with high photon statistics revealed the existence of the neutral iron line. 
ALMA ACA CO ($J=2$--1) observations have shown that SNR W49B interacts with the northwestern molecular cloud (\cite{Sano2021}). 
The present results suggest that the neutral iron line may be occurring in W49B due to interactions with the molecular cloud.
We also identified unknown feature at 8--9 keV. 
In order to elucidate the true origin of the features, future X-ray missions with high spectral resolution such as XRISM are needed.

\begin{ack}
We would like to express our thanks to all of the Suzaku team. 
This work was supported by JST, the establishment of university
fellowships towards the creation of science technology innovation, Grant Number JPMJFS2127 (NS), 
JSPS KAKENHI Grant Number JP21K03615 (MN, SY, KKN), JP20H00174 (SK), and JP21H01121 (SK).
\end{ack}

\appendix
\section*{The spatially resolved spectra}
\begin{figure*}
  \begin{center}
      \FigureFile(16cm,16cm){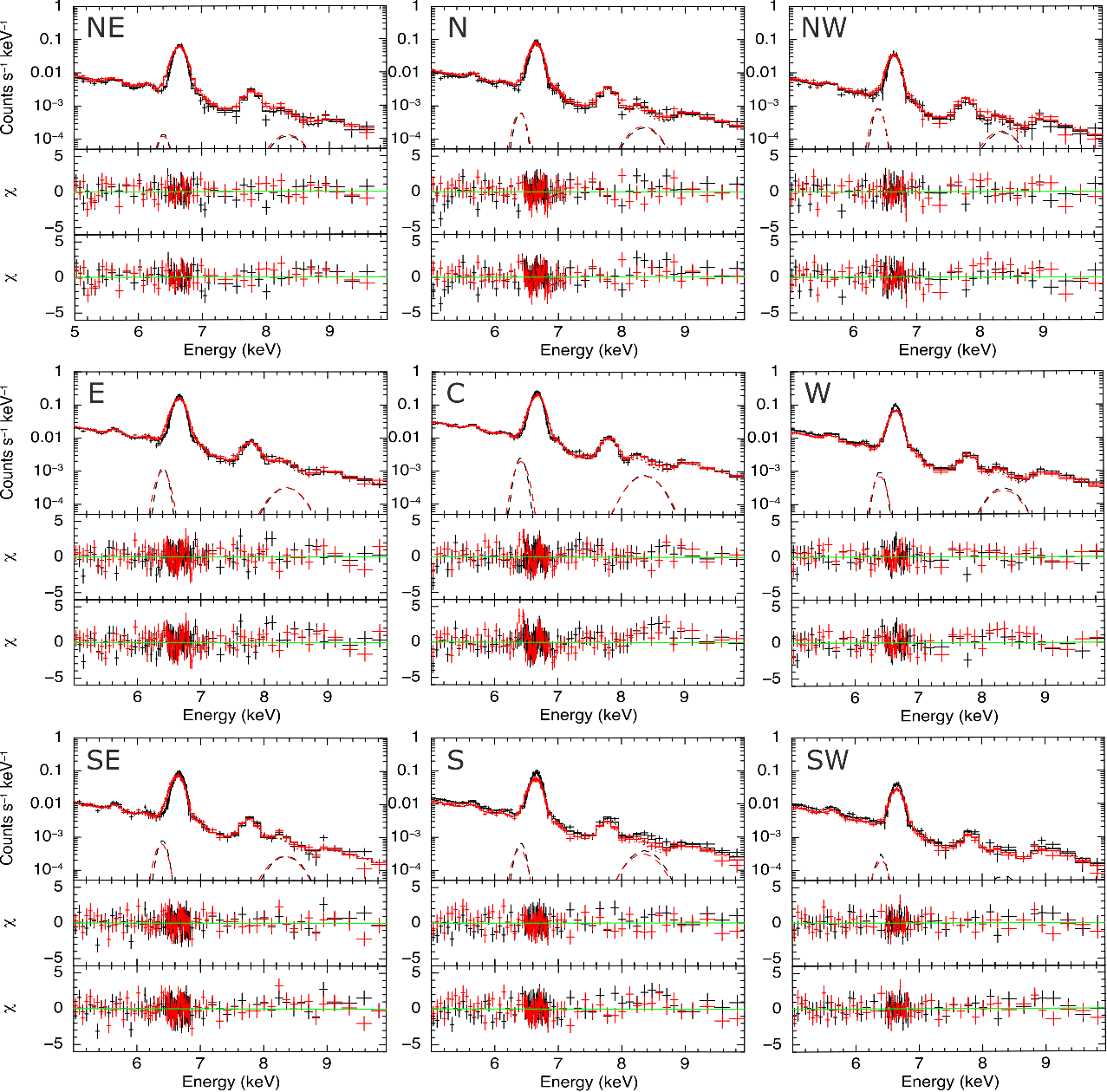}
  \end{center}
  \caption{The 9 region spectra of W49B in the 5--10 keV band (upper panel), residuals from the best-fitting RP$+$2 lines (middle panel), and residuals from the RP model (lower panel).
  The black and red colors show data obtained in 2009 and 2015, respectively.
The histograms show the best-fitting RP$+$2 lines (see table 3).
  }\label{fig:9regspec}
\end{figure*}

 Figure \ref{fig:9regspec} shows the spectra of 9 regions and the best-fitting model.

%%%%%%
% Figures 4
%%
%%%%%%

%%%
% See the manual for the detail.
%%%


\begin{thebibliography}{}
\bibitem[Abdo et al.(2010)]{Abdo2010}
Abdo, A. A., et al. 2010, ApJ, 722, 1303
\bibitem[Anders \& Grevesse(1989)]{Anders1989}
   Anders, E., \& Grevesse, N. 1989, Geochim. Cosmochim. Acta, 53, 197
\bibitem[Balucinska-Church \& McCammon(1992)]{bcmc1992} 
   Balucinska-Church, M., \& McCammon, D. 1992, \apj, 400, 699
   \bibitem[Bambynek et al.(1972)]{Bambynek1972}
   Bambynek, W., Crasemann, B., Fink, R. W., Freund, H.-U., Mark, H., Swift, C. D., Price, R. E., \& Rao, P. V. 1972, Rev. Mod. Phys., 44, 716
\bibitem[Cronin(1999)]{Cronin1999}
   Cronin, J. W. 1999, Rev. Mod. Phys., 71, S165
   \bibitem[Dogiel et al.(2011)]{Dogiel2011}
   Dogiel, V., Chernyshov, D., Koyama, K., Nobukawa, M., \& Cheng, K.-S. 2011, \pasj, 63, 535
   \bibitem[H.E.S.S. Collaboration et al.(2018)]{HESS2018}
   H.E.S.S. Collaboration et al. 2018, A\&A, 612, A5
\bibitem[Hirayama et al.(2019)]{Hirayama2019}
Hirayama, A., Yamauchi, S., Nobukawa, K. K., Nobukawa, M., \& Koyama, K. 2019, \pasj, 71, 37
\bibitem[Hyodo et al.(2008)]{Hyodo2008}
  Hyodo, Y., Tsujimoto, M., Hamaguchi, K., Koyama, K., Kitamoto, S., Maeda, Y., 
  Tsuboi, Y., \& Ezoe, Y. 2008, \pasj, 60, S85
  \bibitem[Ishisaki et al.(2007)]{Ishisaki2007}
  Ishisaki, Y., et al.\ 2007, \pasj, 59, S113
\bibitem[Katsuda et al.(2011)]{Katsuda2011}
   Katsuda, S., et al.\ 2011, \apj, 730, 24
   \bibitem[Katsuda et al.(2012)]{Katsuda2012}
   Katsuda, S., et al.\ 2012, \apj, 756, 49
   \bibitem[Kawasaki et al.(2005)]{Kawasaki2005}
  Kawasaki, M., Ozaki, M., Nagase, F., Inoue, H., \& Petre, R. 2005, \apj, 631, 935
   \bibitem[Kilpatrick et al.(2016)]{Kilpatrick2016}
   Kilpatrick, C. D., Bieging, J. H., \& Rieke, G. H. 2016, \apj, 816, 1
\bibitem[Koyama et al.(2007)]{Koyama2007}
   Koyama, K., et al.\ 2007, \pasj, 59, S23
%   \bibitem[McClure(1966)]{McClure1966}
%   McClure, G. W. 1966, Phys. Rev., 148, 47
\bibitem[Mitsuda et al.(2007)]{Mitsuda2007}
   Mitsuda, K., et al.\ 2007, \pasj, 59, S1
 \bibitem[Mullen et al.(2016)]{Mullen2016}
   Mullen, P. D., Cumbee, R. S., Lyons, D., \& Stancil, P. C. 2016, ApJS, 224, 31
\bibitem[Nakajima et al.(2008)]{Nakajima2008}
   Nakajima, H., et al. 2008, \pasj, 60, S1
   \bibitem[Neronov et al.(2012)]{Neronov2012}
   Neronov, A., Semikoz, D. V., \& Taylor, A. M. 2012, Phys. Rev. Lett., 108, 051105
\bibitem[Nobukawa et al.(2019)]{Nobukawa2019}
   Nobukawa, K. K., Hirayama, A., Shimaguchi, A., Fujita, Y., Nobukawa, M., \& Yamauchi, S. 2019, \pasj, 71, 115
   \bibitem[Nobukawa et al.(2018)]{Nobukawa2018}
   Nobukawa, K. K., et al. 2018, \pasj, 854, 87
\bibitem[Ozawa et al.(2009)]{Ozawa2009}
   Ozawa, M., Koyama, K., Yamaguchi, H., Masai, K., \& Tamagawa, T.
   2009, \apj, 706, L71
\bibitem[Sano et al.(2021)]{Sano2021}
   Sano, H., et al. 2021, \apj, 919, 123
\bibitem[Serlemitsos et al.(2007)]{Serlemitsos2007}
   Serlemitsos, P., et al.\ 2007, \pasj, 59, S9
      \bibitem[Tanaka et al.(2018)]{Tanaka2018}   
   Tanaka, T., et al. 2018, ApJL, 866, L26
      \bibitem[Tatischeff(2003)]{Tatischeff2003}
   Tatischeff, V. 2003, EAS Publications Series, 7, 79
   \bibitem[Tatischeff et al.(2012)]{Tatischeff2012}
   Tatischeff, V., Decourchelle, A., \& Maurin, G. 2012, A\&A, 546, A88
\bibitem[Tawa et al.(2008)]{Tawa2008}
   Tawa, N., et al. 2008, \pasj, 60, S11
  \bibitem[Uchida et al.(2019)]{Uchida2019}
   Uchida, H., Katsuda, S., Tsunemi, H., Mori, K., Gu, L., Cumbee, R. S., Petre, R., \& Tanaka, T. 2019, \apj, 871, 234
\bibitem[Uchiyama et al.(2009)]{Uchiyama2009}
   Uchiyama, H., et al. 2009, \pasj, 61, S9
\bibitem[Wargelin et al.(2005)]{Wargelin2005}
   Wargelin, B. J., Beiersdorfer, P., Neill, P. A., Olson, R. E., \& Scofield, J. H. 2005, \apj, 634, 687
   \bibitem[Wargelin et al.(2008)]{Wargelin2008}
   Wargelin, B. J., Beiersdorder, P., \& Brown, G. V. 2008, Can. J. Phys., 86, 151
\bibitem[Yamaguchi et al.(2018)]{Yamaguchi2018}
   Yamaguchi, H., et al. 2018, \apj, 868, L35
   \bibitem[Zhu et al.(2014)]{Zhu2014}
   Zhu, H., Tian, W. W., \& Zuo, P. 2014, \apj, 793, 95
\end{thebibliography}
\end{document}